\documentclass[12pt]{article}
\usepackage{amsmath,amsfonts,amssymb,graphicx,geometry,authblk,subfig,xcolor,lineno}
\graphicspath{ {Figures/} }
\UseRawInputEncoding
\DeclareMathOperator*{\argmax}{argmax}

\title{Concurrent goal-oriented materials-by-design}
\author{Xingsheng Sun, Burigede Liu, Kaushik Bhattacharya, Michael Ortiz}
\affil{Division of Engineering and Applied Science,\\
    California Institute of Technology,\\
    Pasadena, CA 91125, USA}

\begin{document}
\maketitle

\begin{abstract}

The development of new materials and structures for extreme conditions including impact remains a continuing challenge despite steady advances. Design is currently accomplished using a sequential approach: an optimal material is first developed using the process-structure-properties paradigm, where performance is measured against a blended measure. Then, the structure is optimized while holding the material properties fixed. In this paper, we propose an alternative concurrent and goal-oriented optimization approach where both the material properties and the structure are optimized simultaneously against an overall system-wide performance measure. We develop a non-intrusive, high-performance computational framework based on DAKOTA and GMSH and use it to study the ballistic impact of a double-layer plate of strong AZ31B magnesium alloy and soft polyurea. We show that the proposed concurrent and goal-oriented optimization strategy can provide significant advantage over the traditional sequential optimization approach.

\end{abstract}

\maketitle


\section{Introduction}
\label{sec:intro}

The development of new materials and structures for extreme conditions including impact remains a continuing challenge despite steady advances. The development of new materials follows the process-structure-properties paradigm, where processing is used to control the microstructure in order to obtain the desired properties for a stipulated performance. Performance may involve multiple properties and so-called Ashby maps \cite{ashby1993materials} are used to find the optimal combinations. These maps lead to blended measures of performance such as a strength-to-weight ratio. While initially introduced as an approach to materials selection, this idea has also guided materials-by-design \cite{olson2000designing,alberi20182019,kovachki2021multiscale}, whereby material designs are sought with a particular set of optimal properties. An important aspect of materials-by-design has been the development of materials databases such as AFlow~\cite{taylor2014restful, curtarolo2012aflowlib}, OQMD~\cite{saal2013materials, kirklin2015open} and the Materials Project~\cite{jain2011high, jain2013commentary} using density functional theory, and using data-intensive and machine-learning approaches to search through the databases in order to identify optimal materials.

Materials properties, however, are only one part of the design of complex mechanical systems. Structural and mechanical engineering uses materials as a basis for designing devices and machines that exploit physical principles (balance laws) and material properties (constitutive relations). Structures typically comprise multiple materials that are optimized using a variety of optimization approaches. These approaches include parameter optimization, where a relatively small number of geometric design variables are identified and optimized using gradient-based and non-gradient based methods, as well as more comprehensive topology~\cite{bendsoe2013topology}  and shape~\cite{haftka1986structural} optimization, where the entire structure is designed through optimization. Representative examples include airfoil design for optimal aerodynamical performance~\cite{jeong2005efficient}, structural optimization of bridge design~\cite{sigmund2007topology}, design of energy-absorbing structures in impact~\cite{hou2009crashworthiness}, topology optimization of silicon anode structures~\cite{mitchell2016computational}, and topology optimization of rocket fuel~\cite{kirchdoerfer2019topology}.

Note that there are two optimization steps in this sequential approach. The first step involves optimizing the material properties guided by some overall blended measure of performance relevant to the application, while the second step fixes the material and optimizes the structure. However, different regions of a structure may experience different loading conditions calling for different material properties. This interplay between system-wide structural response and materials properties is not accounted for in the sequential approach. Indeed, a combination of materials, each of which is suboptimal against a blended measure, finely tuned to a specific application, may provide a superior overall structural performance.

In this paper, we propose a goal-oriented approach to simultaneously optimizing materials and structures. To set forth and demonstrate the approach in a practical setting, we specifically focus on the design of protective structures against ballistic impact and penetration. Protection systems typically involve multiple materials, with a strong yet brittle material on the proximal face to the impact and a soft yet ductile material on the distal face, see for example \cite{o2014mechanisms,o2015ballistic,liu2019failure}. We formulate the problem and describe the methodology in Section \ref{sec:problem}. An important step in this direction is to introduce a parametrization of the range of allowable material properties that lends itself to optimization.

We then focus on a double-layer plate made of  AZ31B magnesium (Mg) alloy and polyurea in Section \ref{sec:double}. Due to low mass density and high specific strength, Mg and its alloys have been of significant interest for structural applications, including structures subjected to dynamic loading conditions~\cite{lloyd2016stress}. Polyurea is used as a blast protection coating for strategic installations since it provides exceptional shock and ballistic resistance~\cite{manav2021molecular, cao2021shock}. We describe the range over which the strength and ductility of AZ31B, and the stiffness and failure stress of the polyurea, can vary guided by experimental observations of these materials.

We present the results in Section \ref{sec:results}. Not surprisingly, the double-layer plate performs better than monolithic plates of the same mass. We optimize the double-layer plate using two strategies. The first strategy is sequential, where we take the optimal materials for monolithic plates and optimize the thickness. The second strategy is concurrent and goal-oriented, where we optimize the material and the thickness simultaneously. Remarkably, we find significantly improved performance in the plate optimized using the concurrent and goal-oriented optimization strategy. We conclude with a summary and short discussion in Section \ref{sec:summ}.

\section{Problem formulation and methodology}
\label{sec:problem}

\subsection{General formulation}

We consider a system comprising $M$ materials, each parameterized by a $P$-dimensional vector of parameters $x^m = \{x_p^m\}_{p=1}^P, \ m = 1, \dots, M$. We may vary the parameters of the $m^\text{th}$ material over a set ${\mathcal X}^m$; i.~e., over $x^m \in {\mathcal X}^m$. The geometry of the configuration is parameterized by $G$ geometric parameters $y_g, \ g = 1, \dots, G$. The overall or system objective that we seek to maximize is $\mathcal F(x^m, y_g)$ subject to some constraint ${\mathcal C}(x^m, y_g)$.

In the well-established sequential approach, we also have a blended performance index ${\mathcal B}^m(x^m)$ that describes the performance of the $m^\text{th}$ material. The design problem is formulated as
\begin{equation} \label{eq:seq}
\max_{y_g}  \  {\mathcal F}(\bar{x}^m, y_g) \quad \text{ subject to } \quad {\mathcal C}(\bar{x}^m, y_g) = 0,
\end{equation}
where
\begin{equation} \label{eq:blend}
\bar{x}^m = \argmax_{ x^m\in {\mathcal X}^m} \ {\mathcal B}^m(x^m).
\end{equation}
Thus, we first identify optimal materials according to some blended criterion Eq.~(\ref{eq:blend}) and then use the resulting materials to optimize the geometric parameters Eq.~(\ref{eq:seq}).

In this paper, we propose and describe an alternative concurrent and goal-oriented approach where both the material and geometric parameters are simultaneously chosen to optimize the system objective,
\begin{equation} \label{eq:goal}
\max_{x^m,y_g}  \  {\mathcal F}(x^m, y_g) \quad \text{ subject to } \quad {\mathcal C}(x^m, y_g) = 0 \text{ and }  x^m \in {\mathcal X}^m.
\end{equation}
It is evident from Eq.~(\ref{eq:blend}) and Eq.~(\ref{eq:goal}) that concurrent goal-oriented approach bounds the sequential approach from above. We demonstrate through an example that it can in fact result in significantly superior designs.

\subsection{Ballistic impact of a multi-layer plate}

\begin{figure}
\centering
\includegraphics[width=3.0in]{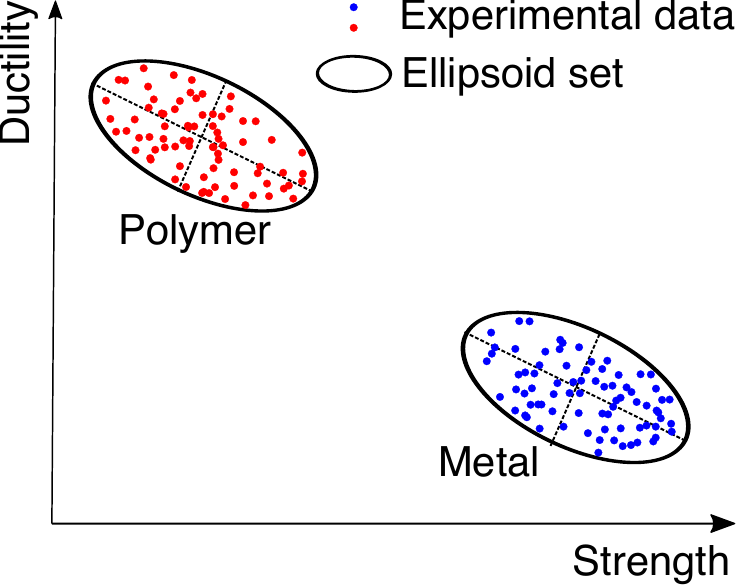}
\caption{Schematic of convex ellipsoidal sets for materials with different strength and ductility.}
\label{fig:ellip}
\end{figure}

For definiteness and for purposes of illustration, we focus on the design of an $M$-layered plate subjected to ballistic impact from a rigid sphere. Each layer is made of a separate material characterized by a set of parameters that describes its mechanical properties. We take the sets ${\mathcal X}^m$ to be ellipsoids encompassing the experimental data~\cite{jiang2011correlation, jiang2013structural}, as shown schematically in Fig.~\ref{fig:ellip}. Specifically, we set
\begin{equation} \label{eq:xm}
{\mathcal X}^m = \left\{ x^m: \big(x^m - x_\text{c}^m \big)^\text{T} W^m
	\big( x^m- x^m_\text{c}  \big) \le 1 \right\}, \quad m=1, \cdots, M ,
\end{equation}
where $W^m$ is a $P\times P$ positive-definite matrix describing the eccentricity of the ellipsoid and $x^m_\text{c}$ is the center. In addition, the geometric design variables are the thicknesses $\{t^m\}$ of the $m^\text{th}$ layer, $G=M, y^g = t^g$. The overall performance metric ${\mathcal F}$ is chosen as the negative of the residual velocity and the design objective is to minimize the residual velocity. The constraint ${\mathcal C}$ is the overall weight of the plate,
\begin{equation}
\mathcal{C}(x^m,t_m) = A\sum_m \rho^mt^m -M_\text{tot}=0 ,
\end{equation}
where $A$ is the area of the plate, $\rho^m$ is the mass density of $m^\text{th}$ layer and $M_\text{tot}$ is the total allowable mass. Finally, the blended performance indicator ${\mathcal B}^m$  is also taken to be ${\mathcal F}$ but for a single material; i.~e., ${\mathcal B}^m(x^m) = {\mathcal F}(x^m, M_\text{tot}/(\rho^m A))$. We specifically focus on a two-layer plate ($M=2$) made of a metal, namely AZ31B, a strong material with high strength but low ductility and a polymer, namely polyurea, a ductile material with high ductility but low strength.

\subsection{Variable transformation}

The bounds on the material variables $x^m$ that we optimize over are not independent because of the constraint $x^m \in {\mathcal X}^m$. We proceed to introduce a change of variables such that the new variables have independent bounds. First, we transform the material parameters into a normalized coordinate space. To this end, we consider the eigen-decomposition of the ellipticity matrices $W^m$
\begin{equation} \label{eq:eigen}
	W^m
	= Q^m \Lambda^m (Q^m)^\text{T}, \quad
	Q^m (Q^m)^\text{T}
	=
	\text{I}, \quad
	m=1,\dots,M,
\end{equation}
where $Q^m\in \mathbb{R}^{P \times P}$ is an orthogonal matrix of the eigenvectors, $\Lambda^m\in \mathbb{R}^{P \times P}$ is a diagonal matrix of the eigenvalues and $\text{I}\in \mathbb{R}^{P \times P}$ is the identity matrix. Based on this decomposition, we introduce the coordinate transformation
\begin{equation} \label{eq:x_q}
	x^m (q^m)
	= x^m_\text{c}
	+
	Q^m	(\Lambda^m)^{-1/2} q^m
\iff
	q^m
	= (\Lambda^m)^{1/2} (Q^m)^\text{T} \big(x^m - x^m_\text{c} \big),
	\quad m=1,\dots,M.
\end{equation}
The domain of the material parameters Eq.~(\ref{eq:xm}) may now be written as $|q^m| \le 1$, i.~e., the normalized parameters must belong to the unit ball. Second, we introduce polar coordinates
\begin{equation} \label{eq:q_r_phi}
\begin{aligned}
	q^m_1 & = r^m \cos(\phi^m_{1}), \\
	q^m_2 & = r^m \sin(\phi^m_{1})  \cos(\phi^m_{2}), \\
	&~ \vdots \\
	q^m_{P-1} & = r^m \sin(\phi^m_{1}) \cdots \sin(\phi^m_{P-2})  \cos(\phi^m_{P-1}), \\
	q^m_{P} & = r^m \sin(\phi^m_{1}) \cdots \sin(\phi^m_{P-1}).
\end{aligned}
\end{equation}
The domain of the material parameters Eq.~(\ref{eq:xm}) is now
\begin{equation} \label{eq:xm2}
    |r^m| \le 1, \quad 0\le \phi^m_{p} < \pi \ (p=1,\dots,P-2), \quad 0\le \phi^m_{P-1} < 2\pi.
\end{equation}
In this representation, the bounds Eq.~(\ref{eq:xm2}) apply independently to each parameter.

\subsection{Optimization problem}

In the new variables just introduced, the \emph{concurrent optimization problem} Eq.~(\ref{eq:goal}) becomes
\begin{equation} \label{eq:opt}
\begin{aligned}
	\max_{\{r\}, \{\phi\}, \{t\}}  \quad & \mathcal{H}\big(\{r\}, \{\phi\}, \{t\}\big), \\
	\text{subject to}  \quad & 0 \le r^m \le 1, \quad m=1,\dots,M, \\
	& 0 \le \phi^m_{p} < \pi, \quad m=1,\dots,M,~ p=1,\dots,P-2, \\
	& 0 \le \phi^m_{P-1} < 2\pi, \quad m=1,\dots,M,  \\
	& A \sum_{m=1}^M\rho^m t^m = M_\text{tot}, \\
	& t^m> 0, \quad m=1,\dots,M,
\end{aligned}
\end{equation}
and the objective function
\begin{equation}
	\mathcal{H}\big(\{r\}, \{\phi\}, \{t\}\big) 
    =
    \mathcal{F}\big(\{x(q(r,\phi))\}, \{t\}\big) ,
\end{equation}
where the functions $x(q)$ and $q(r, \phi)$ are defined in in Eqs.~(\ref{eq:x_q}) and (\ref{eq:q_r_phi}), respectively.

\begin{figure}
\centering
\includegraphics[width=4.5in]{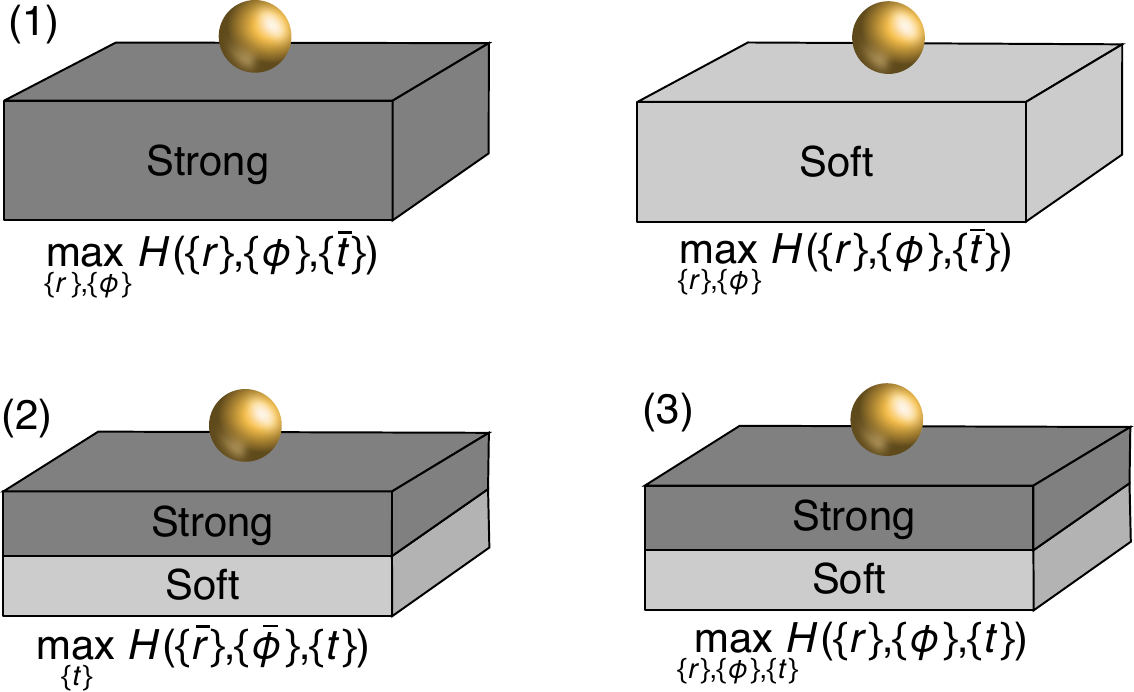}
\caption{Strategies of optimally designing a multi-layered plate subject to high-speed impact. (1) Design over mechanical properties of strong and soft materials separately. (2) Design over structural properties. Steps (1) and (2) form the strategy of sequential design. (3) Concurrent design over mechanical and structural properties.}
\label{fig:case}
\end{figure}

In the \emph{sequential optimization problem}, we first obtain the material parameters $\bar{r}^m, \{\bar{\phi}^m\}$ by maximizing $\mathcal{H}$ with $t^n= 0, \ \forall \ n \ne m$,
\begin{equation} \label{eq:sopt1}
\begin{aligned}
\big(\bar{r}^m, \{\bar{\phi}^m\}\big) =
 \argmax_{r^m, \{\phi^m_p\} } \  & \mathcal{H}\big(\{r\}, \{\phi\}, \{\bar{t}^m\}\big), \\
	\text{subject to}  \quad &  0 \le r^m \le 1 \\
	& 0 \le \phi^m_{p} < \pi,  p=1,\dots,P-2, \\
	& 0 \le \phi^m_{P-1} < 2\pi,  \\
	& \bar{t}^m =  M_\text{tot}/(A\rho^m), \\
	& t^n = 0, \quad n \ne m,
\end{aligned}
\end{equation}
and then solve the optimization problem
\begin{equation} \label{eq:sopt2}
\begin{aligned}
	\max_{ \{t\}} \quad 
    & \mathcal{H}\big(\{\bar{r}\}, \{\bar{\phi} \}, \{t\}\big), \\
	\text{subject to}  \quad & A \sum_{m=1}^M\rho^m t^m = M_\text{tot}, \\
	& t^m> 0, \quad m=1,\dots,M,
\end{aligned}
\end{equation}
to obtain the geometric parameters. An illustration of these two optimization problems is shown in Fig.~\ref{fig:case}.

\subsection{Implementation}

\begin{figure}
\centering
\includegraphics[width=5.0in]{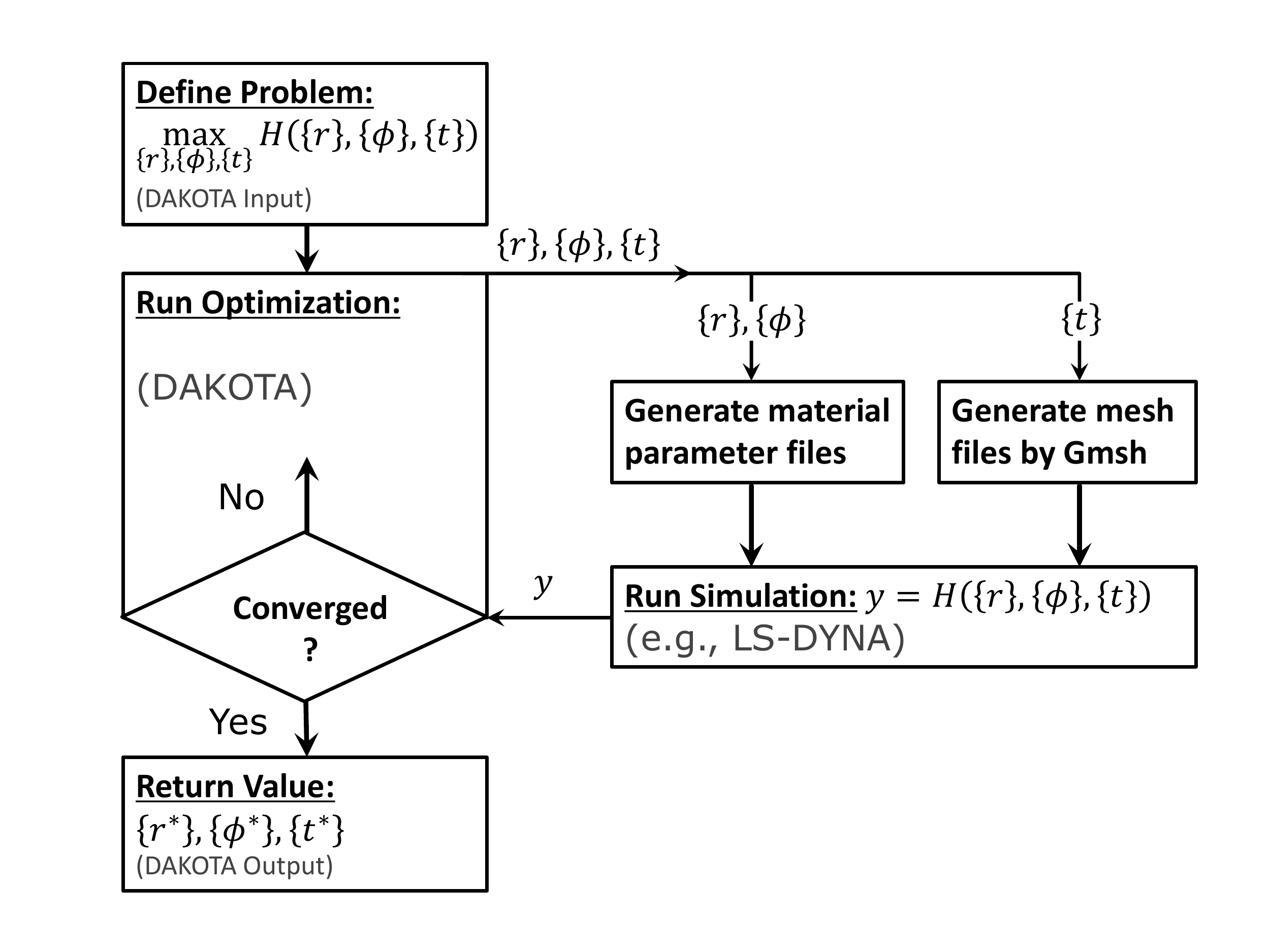}
\caption{Process flow for the concurrent design strategy over mechanical and geometrical properties.}
\label{fig:flowchart}
\end{figure}

Our objective function is the negative of the residual velocity of the projectile. We evaluate the residual velocity using the commercial finite-element analysis program LS-DYNA \cite{hallquist2007ls}. This method of evaluation does not supply the gradients of the objective function, which eliminates gradient-based optimization methods from consideration. Instead, we employ a genetic algorithm (GA)~\cite{mitchell1998introduction, sun2020rigorous, liu2021hierarchical} from DAKOTA Version $6.12$ software package~\cite{adams2020dakota} of the Sandia National Laboratories. As a global and derivative-free optimization method, GAs offer great flexibility in complex applications such as considered here. Another advantage of GAs is their high degree of concurrency, since the fitness of individuals can be evaluated independently across multiple processors. In all the GA calculations, we choose a fixed population size of $64$. The crossover rate and mutation rate are fixed at $0.8$ and $0.1$. In addition, we have integrated the Gmsh Version $4.5.4$ software package~\cite{geuzaine2009gmsh} into the framework, since the optimization requires evaluation of different structural parameters and hence the generation of finite element meshes on demand. A flowchart of the concurrent goal-oriented design strategy is shown in Fig.~\ref{fig:flowchart}.

\section{Ballistic impact of double-layered plates} \label{sec:double}

We assess the sequential and concurrent design strategies by means of a test case consisting of a double-layered plate of AZ31B magnesium (Mg) alloy and polyurea.


\subsection{AZ31B Mg alloy}

We assume that AZ31B obeys an appropriately calibrated Johnson-Cook (JC) plasticity model~\cite{johnson1983constitutive},
\begin{equation}
    \sigma \big( \epsilon_p, \dot{\epsilon}_p, T \big)
    =
    \big[A + B \epsilon_p^n \big]
    \big[1 + C \ln \dot{\epsilon}_p^* \big]
    \big[1 - {T ^*} ^m \big],
\label{eq:johnsoncook}
\end{equation}
where $\sigma$ is the true Mises stress, $\epsilon_p$ is the equivalent plastic strain, $\dot{\epsilon}_p$ is the plastic strain rate, and $T$ is the temperature. The normalized plastic strain rate $\dot{\epsilon}_p^*$ is defined as
\begin{equation}
    \dot{\epsilon}_p^* := \frac{\dot{\epsilon}_p}{\dot{\epsilon}_{p0}},
\label{eq:estar}
\end{equation}
where $\dot{\epsilon}_{p0}$ is a reference strain rate. The model also uses the normalized temperature
\begin{equation}
    T^* := \frac{T - T_0}{T_m - T_0},
\label{eq:tstar}
\end{equation}
where $T_0$ is a reference temperature and $T_m$ is the melting temperature. The model parameters are: $A$, the yield stress; $B$, the strain-hardening modulus; $n$, the strain-hardening exponent; $C$, the strengthening coefficient of strain rate; and $m$, the thermal-softening exponent.

We further assume that the failure behavior of AZ31B is well-described by the Johnson-Cook fracture model~\cite{johnson1985fracture}. Specifically, the damage of an element is defined on a cumulative damage parameter
\begin{equation}
    D
    =
    \sum\frac{\Delta \epsilon}{ \epsilon^f},
\label{eq:damageparam}
\end{equation}
where the summation is conducted over time steps and $\Delta \epsilon$ is the plastic strain increment in each time step. $ \epsilon^f$ denotes the strain at fracture which is given by
\begin{equation}
    \epsilon^f
    =
    \big[D_1 + D_2 \exp(D_3 \sigma^*)\big]
    \big[1 + D_4 \ln \dot{\epsilon_p}^* \big]
    \big[1 + D_5 T^* \big],
\label{eq:fracturestrain}
\end{equation}
where $D_1$, $D_2$, $D_3$, $D_4$ and $D_5$ are material damage constants. ${\epsilon_p}^*$ and $T^*$ are defined in Eqs.~(\ref{eq:estar}) and (\ref{eq:tstar}), respectively. $\sigma^*$ is the ratio of the pressure $p$ divided by the von-Mises equivalent stress, i.~e.,
$
    \sigma^*
    =
    p\left(\big[ (\sigma_1-\sigma_2)^2 + (\sigma_2-\sigma_3)^2 + (\sigma_3-\sigma_1)^2 \big]/2\right)^{-1/2},
$
where $\sigma_1$, $\sigma_2$ and $\sigma_3$ are the principle stresses. Based on this damage model, fracture takes place when the damage parameter $D$ reaches the value of $1$. Finally, the equation-of-state, which controls the volumetric response of the material, is assumed to be of the Gruneisen type.

\subsection{Polyurea}

Polyurea is assumed to obey the Mooney-Rivlin (MR) model~\cite{ogden1997non}, which describes polyurea as a isotropic hyperelastic rubber-like material with a strain energy density depending on the invariants of the left Cauchy-Green deformation tensor. Specifically, the strain energy density is of the form
\begin{equation}
    W(\bar{I}_1, \bar{I}_1, J)
    = \sum^n_{p,q=0} C_{pq} (\bar{I}_1-3)^p (\bar{I}_2-3)^q + W_\text{H}(J),
\label{eq:mooney}
\end{equation}
with
\begin{equation}
    \bar{I}_1 = I_1 I_3^{-1/3},~
    \bar{I}_2 = I_2 I_3^{-2/3},~
    J = I_3^{1/2},
\end{equation}
where $I_1$, $I_2$ and $I_3$ are the invariants of the left Cauchy-Green deformation tensor. $W_\text{H}(J)$ characterizes the contribution of hydrostatic work to the strain energy. In this work, we adopt a six-term version of Eq.~(\ref{eq:mooney}) and the material parameters involved are $C_{10}$, $C_{01}$, $C_{11}$, $C_{20}$, $C_{02}$ and $C_{30}$.

Strain rate effects are accounted by assuming linear viscoelasticity in the form of a convolution integral
\begin{equation}
    \sigma_\text{a} = \int_0^t g(t-\tau) \frac{\text{d}\epsilon}{\text{d}\tau}\text{d}\tau,
\end{equation}
where $\sigma_\text{a}$ is the stress tensor that is added to the stress tensor determined from the strain energy functional, $\epsilon$ denotes the strain tensor and $g(t)$ is the relaxation modulus. We additionally express $g(t)$ as a Prony series
\begin{equation}
    g(t) = \sum_{i=1}^n g_i e^{-\beta_i t},
\end{equation}
where $g_i$ and $\beta_i$ denote shear moduli and decay constants, respectively. Finally, the attainment of a critical value of the maximum principal stress $\sigma_\text{max}$ is adopted as the failure criterion for polyurea.

\subsubsection{Constraints on mechanical properties}

\begin{figure}[t]
\centering
\subfloat[ ]{\includegraphics[trim=0.5in 0.0in 0.5in 0.0in,clip,height=2.2in]{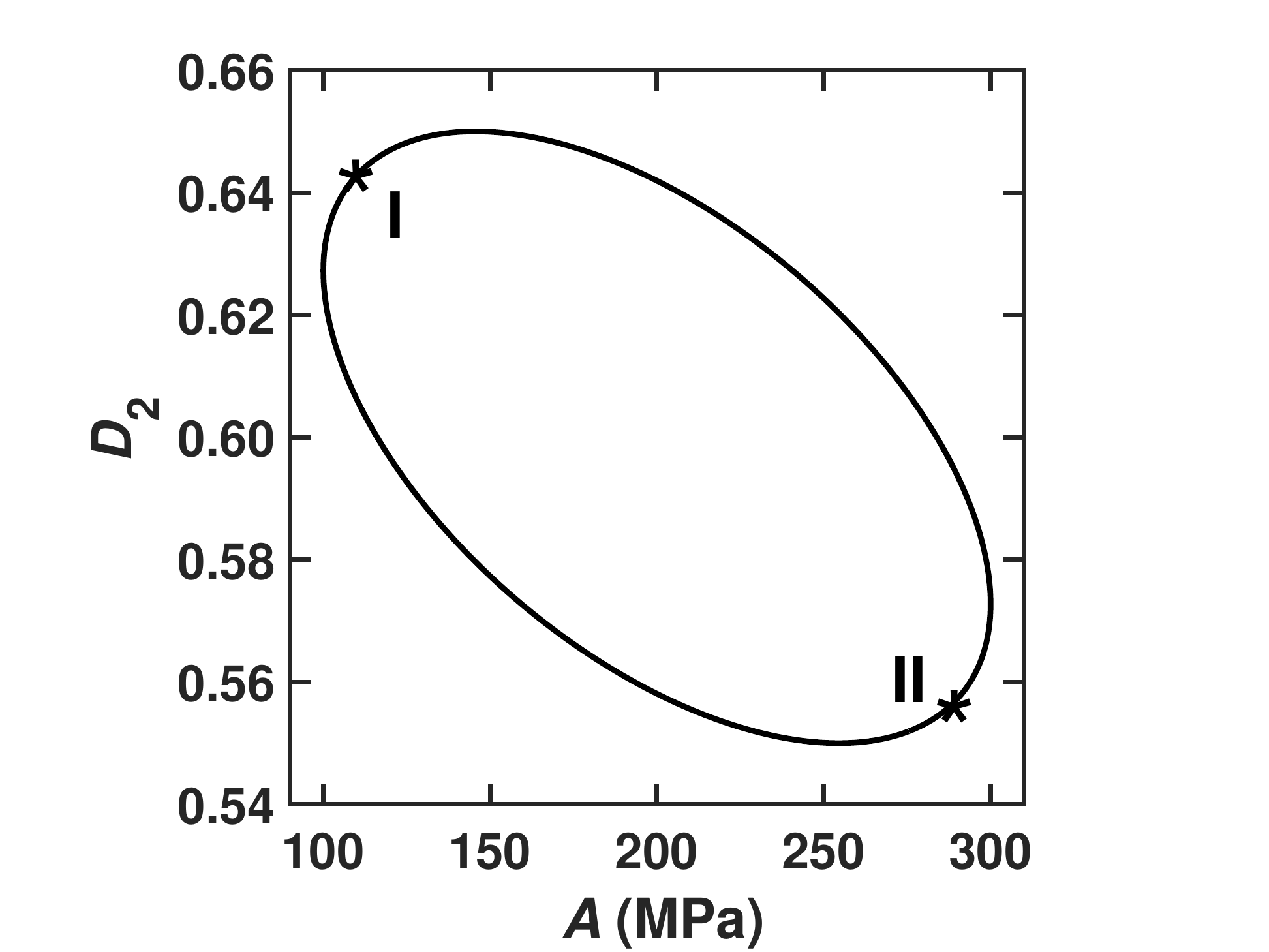}}
\subfloat[ ]{\includegraphics[trim=0.5in 0.0in 0.5in 0.0in,clip,height=2.2in]{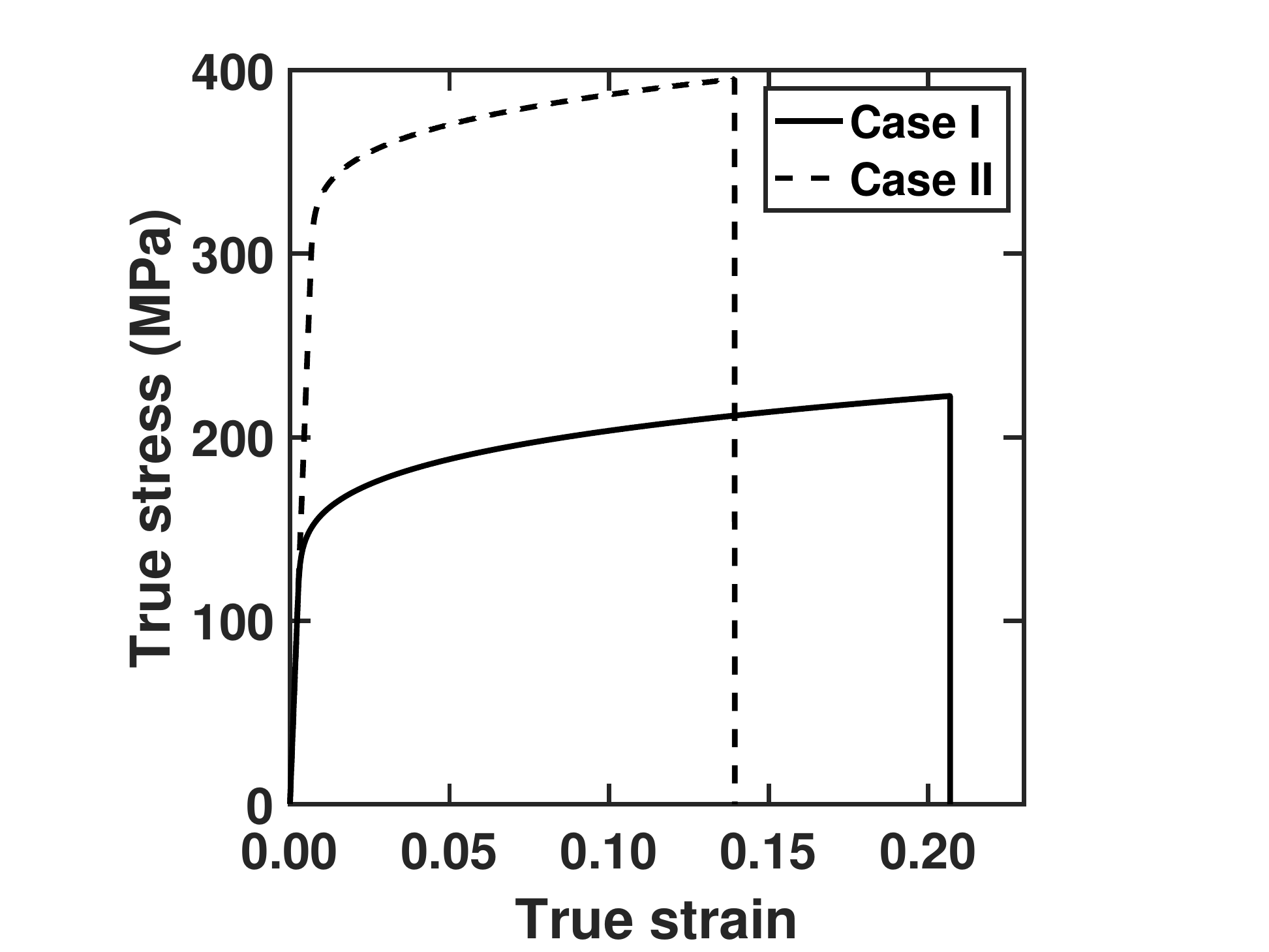}}\\
\subfloat[ ]{\includegraphics[trim=0.5in 0.0in 0.5in 0.0in,clip,height=2.2in]{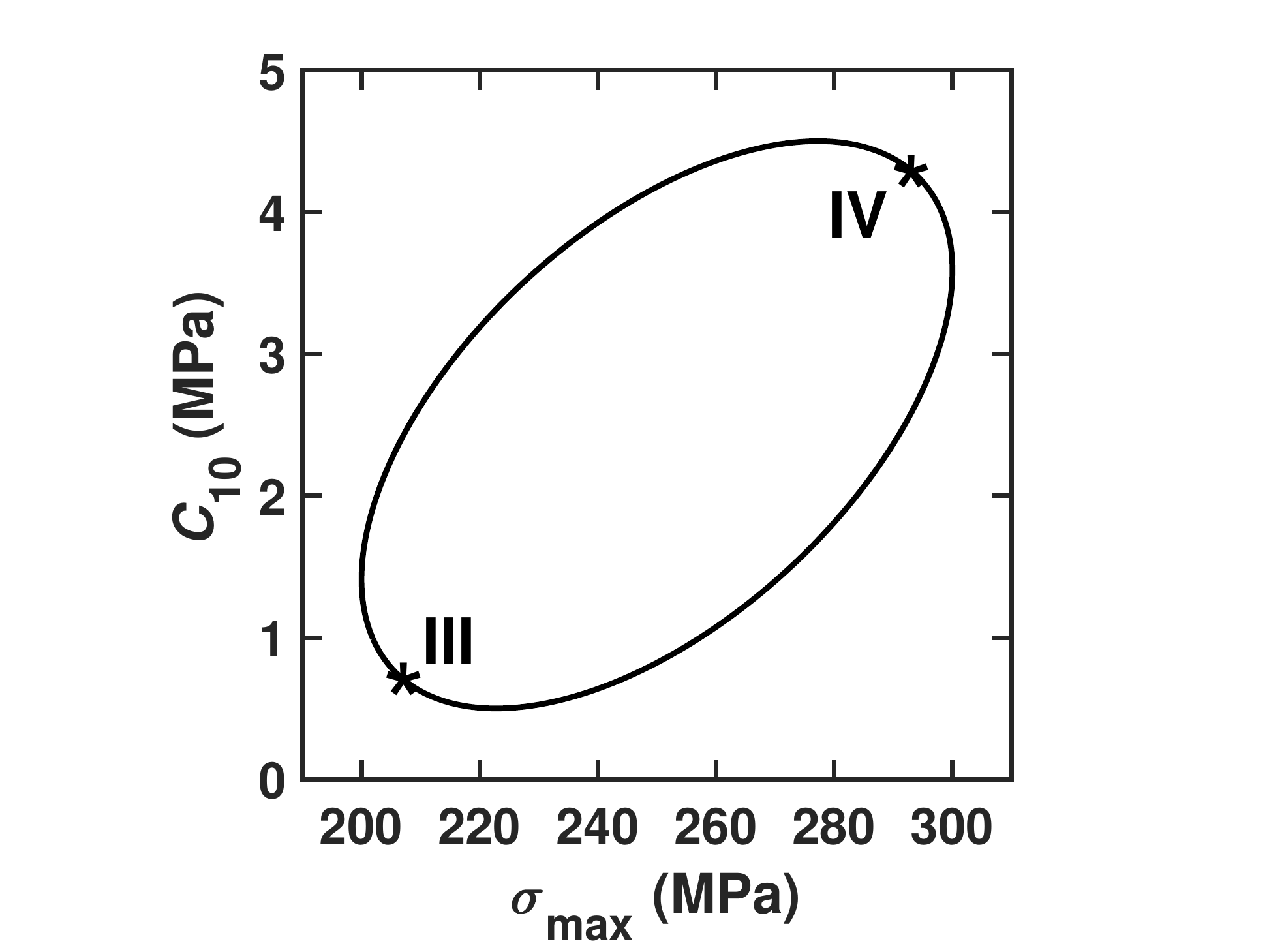}}
\subfloat[ ]{\includegraphics[trim=0.5in 0.0in 0.5in 0.0in,clip,height=2.2in]{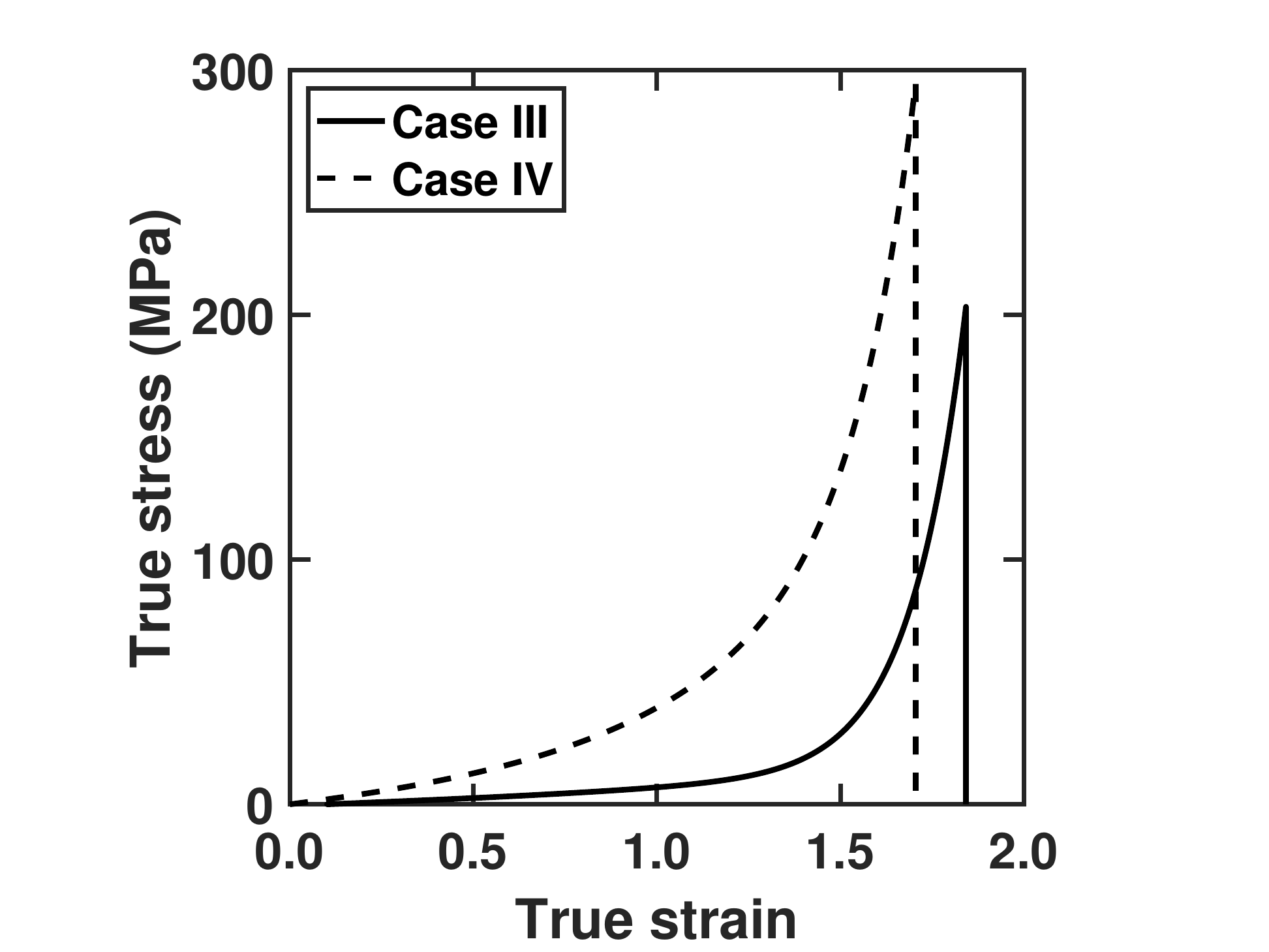}}
\caption{The range of material parameters and representative stress-strain curves for (a,b) AZ31B and (c,d) Polyurea. The stress-strain curves in (b) (respectively (d)) correspond to point marked in (a) (respectively (c)).}
\label{fig:mat}
\end{figure}

We regard  $x^1 = (A, D_2)$ and $x^2 = (\sigma_\text{max}, C_{10})$ as the design variables characterizing the mechanical properties of AZ31B and polyurea, respectively. In practice, the ranges of material parameters are derived from specific data sources, e.~g., experiments or sub-grid simulations. In our calculations, in view of the paucity of the available archival data we simply assume that these design variables are constrained by the convex ellipsoidal sets depicted in Fig.~\ref{fig:mat}. The remaining values of the material parameters used in the calculations are tabulated in Tables~\ref{tab:paramAZ31B}-\ref{tab:paramprojectile}. Representative  stress-strain relations for uniaxial tension are also shown in Fig.~\ref{fig:mat}. We note that the strength of the materials increases while the ductility decreases from Case I to II in AZ31B and from Case III to IV in polyurea. Thus, the ellipsoidal material domains capture the negative correlation between strength and ductility in both materials.

\begin{table}
\centering
\caption{Fixed material parameters of AZ31B used in the ballistic problem.}
\begin{tabular}{l l l l}
\hline
\hline
Plate (AZ31B)              &     Value               & Unit     & Source               \\
\hline
Mass density                    &     $1.77$              & g/$\text{cm}^3$       \\
Young's modulus                 &     $45.0$              & GPa                    \\
Poisson's ratio                 &     $0.35$              & -                         \\
Specific heat                 & $1.75$ &  J/(K$\cdot$g) & \cite{lee2013thermal}  \\
Taylor-Quinney factor  & $0.6$   & -                      & \cite{kingstedt2019conversion}  \\
Spall strength              & $1.5$    & GPa                 & \cite{farbaniec2016microstructural}  \\
Gruneisen intercept     & $4520.0$ & m/s  & \cite{feng2017numerical}  \\
Gruneisen gamma        & $1.54$    & -       & \cite{feng2017numerical}  \\
Gruneisen slope $S_1$ & $1.242$  & -       & \cite{feng2017numerical}  \\
Reference strain rate      &  $0.001$  & $\text{s}^{-1}$ & \cite{hasenpouth2010tensile}  \\
Reference temperature   &  $298.0$  & K                      & \cite{hasenpouth2010tensile}  \\
Reference melt. temp. &  $905.0$  & K                      & \cite{hasenpouth2010tensile}  \\
JC parameter $B$ &     $168.346$  & MPa &\cite{hasenpouth2010tensile}  \\
JC parameter $n$ &     $0.242$      & -     &\cite{hasenpouth2010tensile}  \\
JC parameter $C$ &     $0.013$      & -     &\cite{hasenpouth2010tensile}  \\
JC parameter $m$ &     $1.55$       & -     &\cite{hasenpouth2010tensile}  \\
JC parameter $D_1$ &  $-0.35$     & -     &\cite{feng2014constitutive}  \\
JC parameter $D_3$ &  $-0.4537$ & -     &\cite{feng2014constitutive}  \\
JC parameter $D_4$ &  $-0.4738$ & -     &\cite{feng2014constitutive}  \\
JC parameter $D_5$ &  $7.2$         & -     &\cite{feng2014constitutive}  \\
\hline
\end{tabular}
\label{tab:paramAZ31B}
\end{table}

\begin{table}
\centering
\caption{Fixed material parameters of polyurea used in the ballistic problem.}
\begin{tabular}{l l l l}
\hline
\hline
Plate (polyurea) &     Value               & Unit     & Source               \\
\hline
Mass density                    &     $1.07$              & g/$\text{cm}^3$       \\
Equil. shear modulus                 &     $0.0224$              & GPa                    \\
Poisson's ratio                 &     $0.485$              & -                         \\
MR parameter $C_{01}$         &     $0.7$              & MPa & \cite{xue2010penetration}  \\
MR parameter $C_{11}$         &     $-0.03$              & MPa & \cite{xue2010penetration}  \\
MR parameter $C_{20}$         &     $-0.02$              & MPa & \cite{xue2010penetration}  \\
MR parameter $C_{02}$         &     $0.001$              & MPa & \cite{xue2010penetration}  \\
MR parameter $C_{30}$         &     $0.002$              & MPa  & \cite{xue2010penetration}  \\
Prony parameter $g_1$         &     $0.0189$              & GPa &  \cite{amirkhizi2006experimentally}  \\
Prony parameter $\beta_1$   &     $2.158$              & $\text{s}^{-1}$ &  \cite{amirkhizi2006experimentally}  \\
Prony parameter $g_2$         &     $0.0378$              & GPa   &  \cite{amirkhizi2006experimentally}  \\
Prony parameter $\beta_2$   &     $1.5608\times 10^4$   & $\text{s}^{-1}$ &  \cite{amirkhizi2006experimentally}  \\
Prony parameter $g_3$         &     $0.0805$              & GPa    &  \cite{amirkhizi2006experimentally}  \\
Prony parameter $\beta_3$   &     $8.5985\times 10^6$   & $\text{s}^{-1}$   &  \cite{amirkhizi2006experimentally}  \\
Prony parameter $g_4$         &     $0.0973$              & GPa   &  \cite{amirkhizi2006experimentally}  \\
Prony parameter $\beta_4$  &     $1.3659\times 10^9$   & $\text{s}^{-1}$   &  \cite{amirkhizi2006experimentally}  \\
\hline
\end{tabular}
\label{tab:parampolyurea}
\end{table}

\begin{table}
\centering
\caption{Fixed material parameters of the projectile used in the ballistic problem.}
\begin{tabular}{l l l}
\hline
\hline
Projectile (steel)           &     Value                &  Unit           \\
\hline
Mass density                   &     $7.83$              & g/$\text{cm}^3$       \\
Young's modulus                &     $210.0$               & GPa                   \\
Poisson's ratio                &     $0.30$               &  -                      \\
\hline
\end{tabular}
\label{tab:paramprojectile}
\end{table}

\subsection{Forward solver}

\begin{figure}
\centering
\includegraphics[width=4.0in]{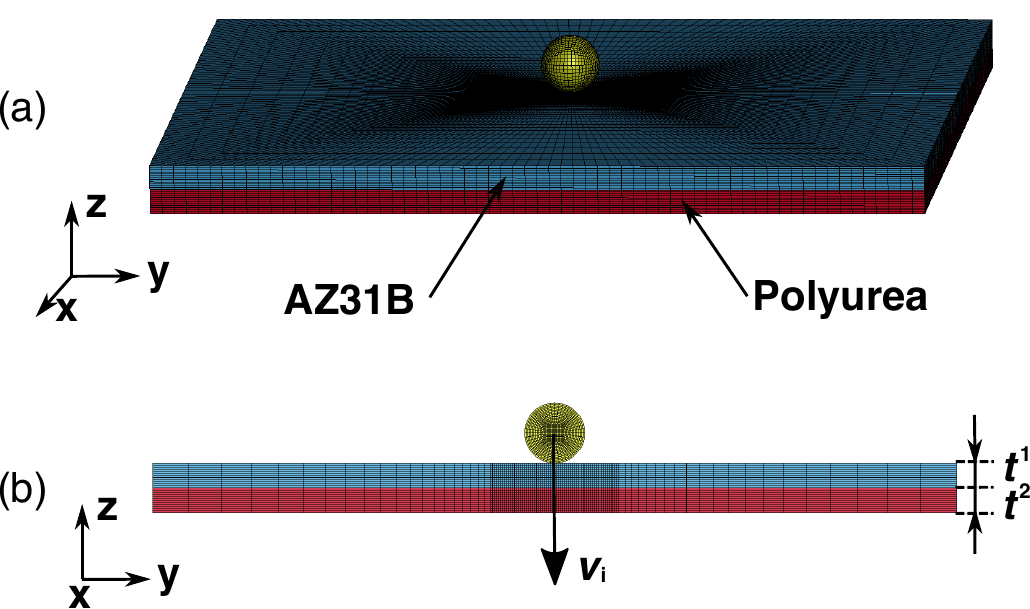}
\caption{Schematic of the computational model. (a) Perspective view of the projectile/plate system. (b) Mid $y$-$z$ cross-section showing impact velocity $v_\text{i}$ and layer thicknesses $t^1$ and $t^2$.}
\label{fig:initial}
\end{figure}

For a given realization of the material parameters, the residual velocity $v_\text{r}$ of the projectile after penetrating the plate is computed using the explicit dynamics solver available in the commercial finite-element analysis software package LS-DYNA \cite{hallquist2007ls}. The initial conditions of the computational model are shown in Fig.~\ref{fig:initial}(a). Fig.~\ref{fig:initial}(b) shows a cross-sectional view of the system before deformation with labeled variables of interest. The plate is square of size $10$~cm in all calculations and is free standing. The projectile is spherical with diameter $0.747$~cm.
The impact velocity is $v_\text{i}=400$~m/s at normal incidence. We take the total mass of the plate $M_\text{tot}$ as free or operating parameter for purposes of parametric studies. The projectile is resolved using $19,208$ elements, while the number of elements for the plate depends on the thicknesses of layers, e.~g., $221,184$ elements for $t^1=t^2=0.31$~cm. All the elements are linear hexahedra with one-point integration and hourglass control. The element size is refined in the impact region of the plate. The interface between layers is assumed to be conforming. The calculations terminate when full perforation of the plate has been achieved. The time-step size is adaptive and determined by the size of elements. The calculations are adiabatic with the initial temperature set to room temperature.

\section{Results} \label{sec:results}

\subsection{Comparison between design strategies}

We assess the sequential and concurrent design strategies under the operating condition $M_\text{tot}=88.5$~g. The optimal design parameters are shown in  Table~\ref{tab:result} together with the performance both in terms of the residual velocity and the total dissipated energy of each plate, computed as
\begin{equation}
    E_\text{d} = \frac{1}{2} M_\text{p}\big(v_\text{i}^2-v_\text{r}^2\big) ,
\label{eq:diss}
\end{equation}
where $M_\text{p}$ is the mass of the projectile. It is evident from the table that the rank-ordering of ballistic performance is
\begin{equation}
\text{concurrent} > \text{sequential} > \text{AZ31B}>\text{polyurea}.
\end{equation}
The double-layered plates resulting from both design strategies perform better than the monolithic counterparts under high speed impact. In terms of energy dissipated, the concurrent design improves the ballistic performance of the plate by $33.0\%$ over monolithic AZ31B, by $48.5\%$ over monolithic polyurea, and by $9.6\%$ relative to the sequential design.

\begin{figure}
\centering
\includegraphics[width=3.0in]{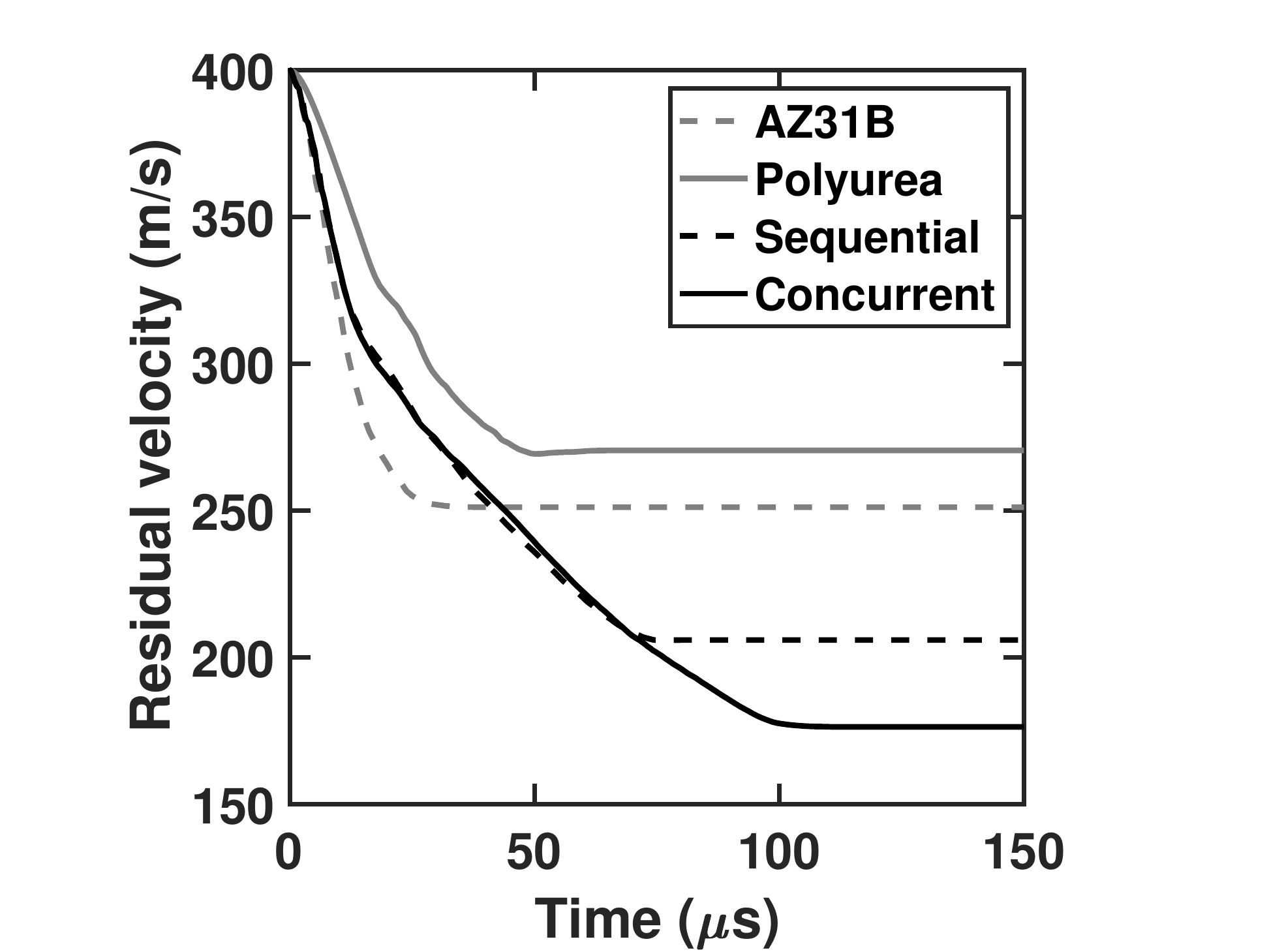}
\caption{Time history of residual velocity at the optimal mechanical and structural properties.}
\label{fig:residual}
\end{figure}

Fig.~\ref{fig:residual} shows the velocity of the projectile as a function of time for the four cases. Initially, from $0$ to $15~\mu$s, the velocity in both concurrent and sequential designs exhibits a history similar to that of monolithic AZ31B. This similarity is not surprising, since the material properties of AZ31B are themselves similar in all three cases (Table~\ref{tab:result}). Subsequently, from $15$ to $75~\mu$s, the general trend in the velocity of both double-layered plates resembles that of the monolithic polyurea. However, after roughly $75~\mu$s, the velocity in the concurrent design continues to decrease, with failure polyurea layer the rate-controlling mechanism. It is this final stage of penetration that sets the difference between the sequential and concurrent strategies.

\begin{table}
\centering
\caption{Optimal design parameters and performance in the case of $M_\text{tot}=88.5$~g.}
\begin{tabular}{l l l l l}
\hline
\hline
\multicolumn{5}{c}{Optimal parameters}\\
\hline
                                             & AZ31B & Polyurea & Sequential & Concurrent         \\
\hline

$A$~(MPa)                            & $294.66$ & -             & $294.66$ & $296.25$ \\
$D_2$                                   & $0.5871$  & -            & $0.5871$ & $0.5738$ \\
$\sigma_\text{max}$~(MPa) &  -             & $290.03$ & $290.03$ & $289.80$  \\
$C_{10}$~(MPa)                    &  -             & $4.3356$ & $4.3356$ & $2.4379$ \\
$t_1$~(cm)                          & 0.5               & -              & $0.3287$ & $0.3538$ \\
$t_2$~(cm)                          & -               & 0.8045              & $0.2757$ & $0.2353$\\
\hline
\hline
\multicolumn{5}{c}{Performance}\\
\hline
                                             & AZ31B & Polyurea & Sequential & Concurrent         \\
\hline
$v_\text{r}$~(m/s)               & $251.17$ & $270.54$ & $205.95$ & $176.36$\\
$E_\text{d}$~(J)                   & $10.58$ & $9.47$ & $13.72$ & $14.89$ \\
\hline
\hline
\end{tabular}
\label{tab:result}
\end{table}

\begin{figure}
\centering
\includegraphics[width=4.0in]{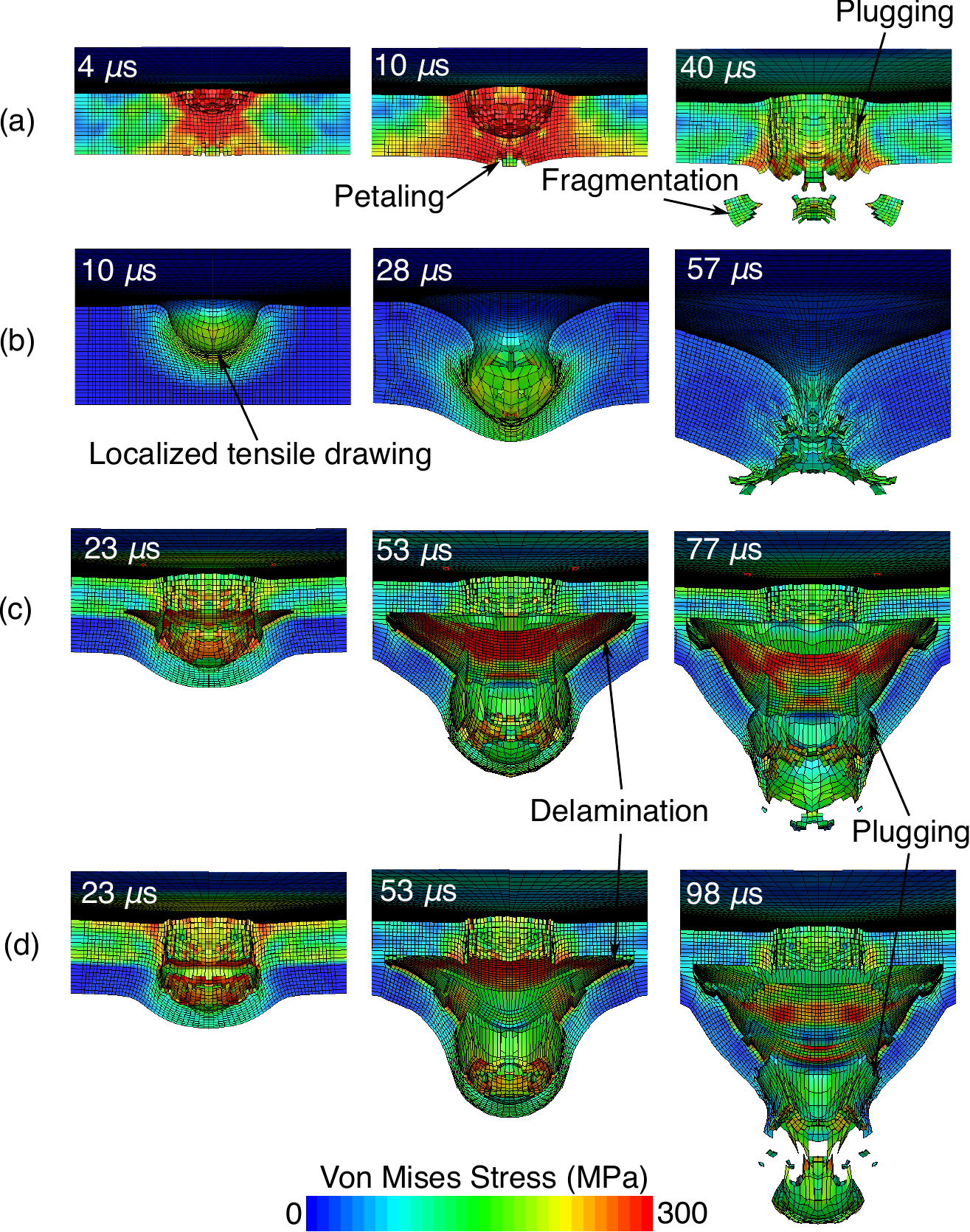}
\caption{Von Mises stress on the cross-section of the impact region at different time instances. (a) Monolithic AZ31B plate. (b) Monolithic polyurea plate. (c) Double-layered plate by sequential design. (d) Double-layered plate by concurrent design. The projectile is removed for the sake of clarity.}
\label{fig:process}
\end{figure}

To understand this velocity history, we examine the material failure mechanisms involved in the design results. Fig.~\ref{fig:process} shows snapshots of the von Mises effective stress on the middle cross-section of the four plates at three different time instances. Fig.~\ref{fig:process}(a) shows that the AZ31B plate is perforated mainly by a conventional shear plugging mechanism. The impact energy is dissipated into plastic work around the shear band that demarcates the perforation area. In addition to plugging, petaling occurs at $10~\mu\text{s}$ near the backface of the AZ31B plate due to the large tensile hoop stresses induced by the bending deformation of the plate. The calculation also reveals material fragmentation at $40~\mu\text{s}$, which is typical of brittle materials such as AZ31B Mg alloys. In contrast, in the polyurea plate Fig.~\ref{fig:process}(b) shows that material failure takes place immediately under the projectile through predominantly a localized tensile drawing mechanism~\cite{mohagheghian2015impact}. 

The combination of AZ31B and polyurea on the proximal and distal face of the plate can significantly improve the ballistic performance relative to that of monolithic designs. The basis for the superior performance can again be gleaned from an examination of the failure mechanisms involved in the double-layered plates. Specifically, Figs.~\ref{fig:process}(c) and (d) reveal that the AZ31B layer in both designs undergoes shear plugging. Significant petaling does not occur due to the confining effect of the polyurea layer on the back. In addition, the failure mode of polyurea also changes compared to the monolithic design. Thus, shear plugging operates in addition to tensile drawing, since the hard AZ31B front layer constrains the bending of the back polyurea layer. The plugging mechanism is activated when the shear stress reaches the shear strength of the polyurea. In addition, delamination is also observed to occur between the AZ31B and polyurea layers.

The superior performance of concurrent over sequential designs owes mainly to the difference in the optimized mechanical properties of the polyurea layer and the attendant differences in material failure mechanisms. Thus, from Table~\ref{tab:result} it is seen that the parameter $C_{10}$ in the concurrent design is smaller than that in the sequential design. Since the ductility of the polyurea decreases monotonically with increasing $C_{10}$, cf.~Fig.~\ref{fig:mat}, the polyurea in the concurrent design is more ductile. As a result, delamination occurs later in the concurrent design relative to the sequential design, as shown in the snapshots at $23~\mu$s in Figs.~\ref{fig:process}(c) and (d). Further, the projectile fully penetrates the plate at $77~\mu$s in the sequential design but only at $98~\mu$s in the concurrent design. Thus, the plugging mechanism operates longer in the concurrent design and hence dissipates more impact energy.

In the sequential design, the polyurea properties are determined by a consideration of monolithic plates and optimizing a blended measure of strength and ductility. However, in the double-layer plate, the AZ31B provides the strength and the role of the polyurea is to provide the ductility. The concurrent design strategy therefore sacrifices strength in favor of ductility in the polyurea, a trade-off that is outside the scope of the sequential design strategy.

\subsection{Parametric study}

\begin{figure}
\centering
\subfloat[ ]{\includegraphics[width=6.0in]{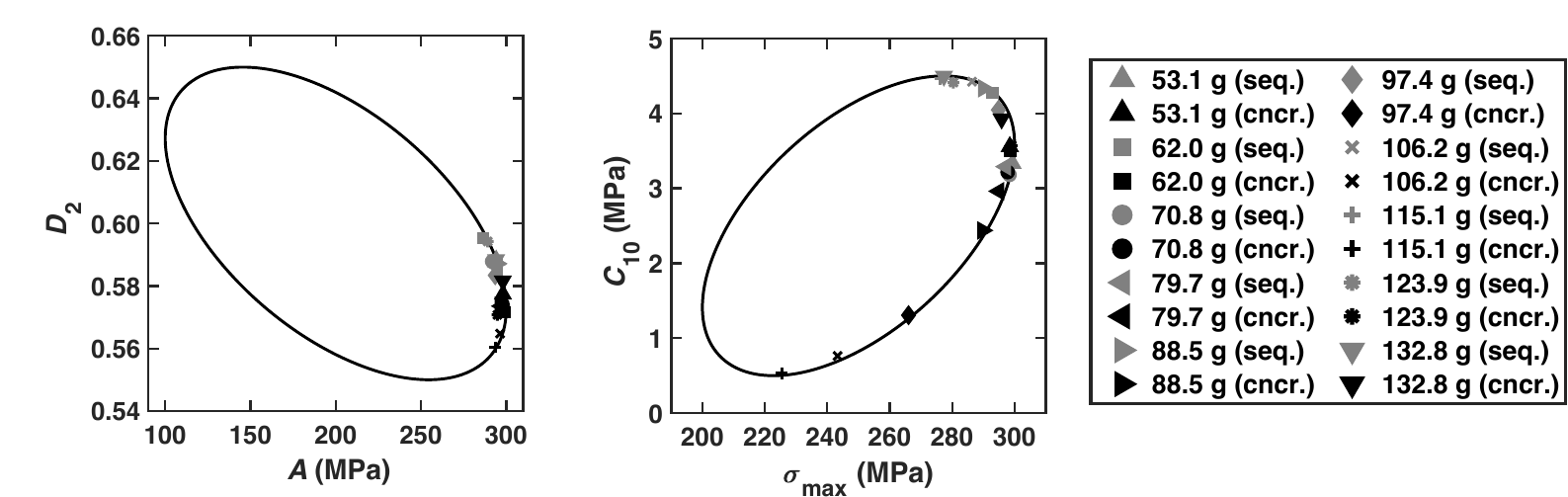}}\\
\subfloat[ ]{\includegraphics[trim=0.9in 0 1.1in 0, clip, width=2.0in]{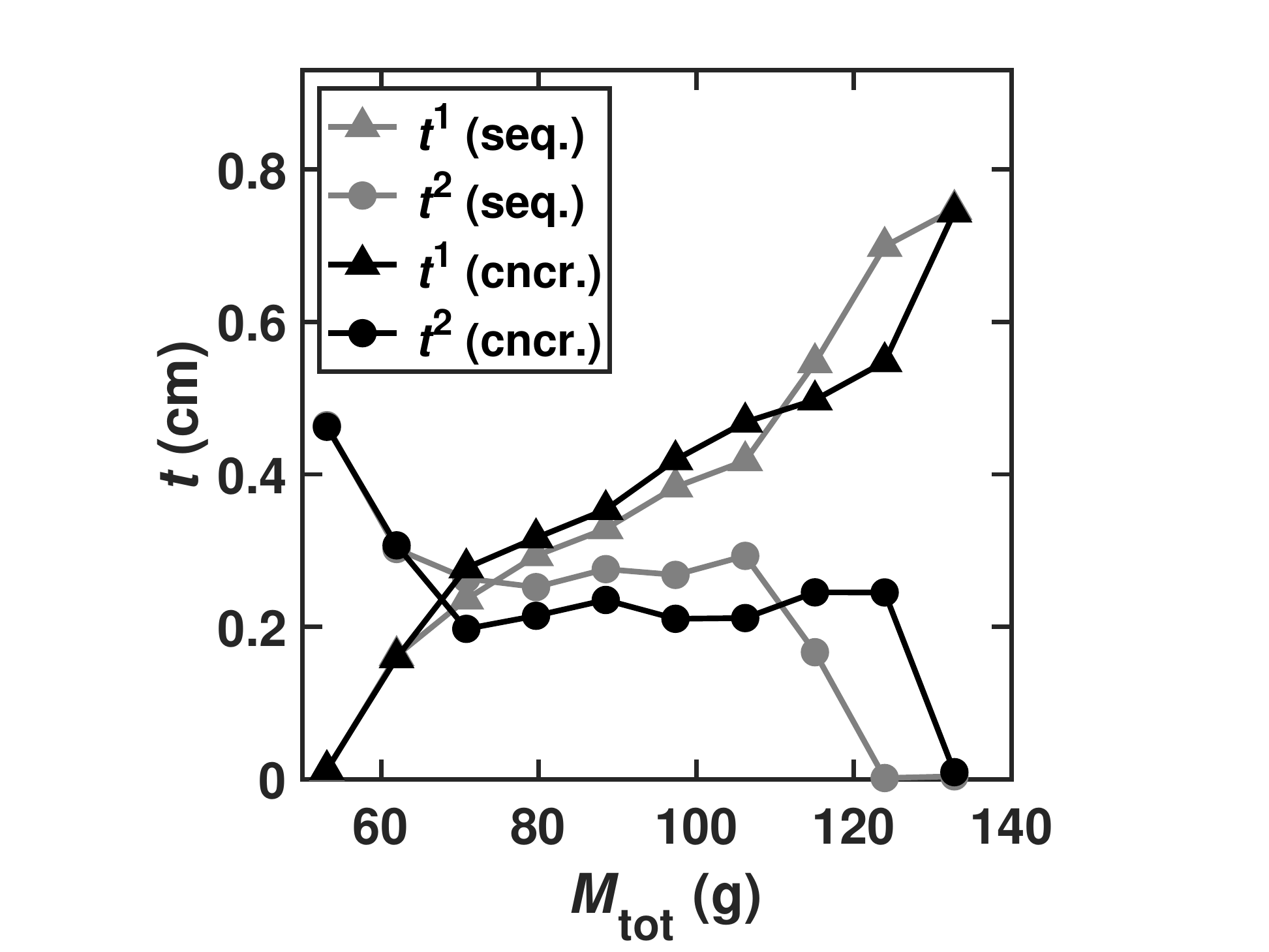}}
\subfloat[ ]{\includegraphics[trim=0.7in 0 1.3in 0, clip, width=2.0in]{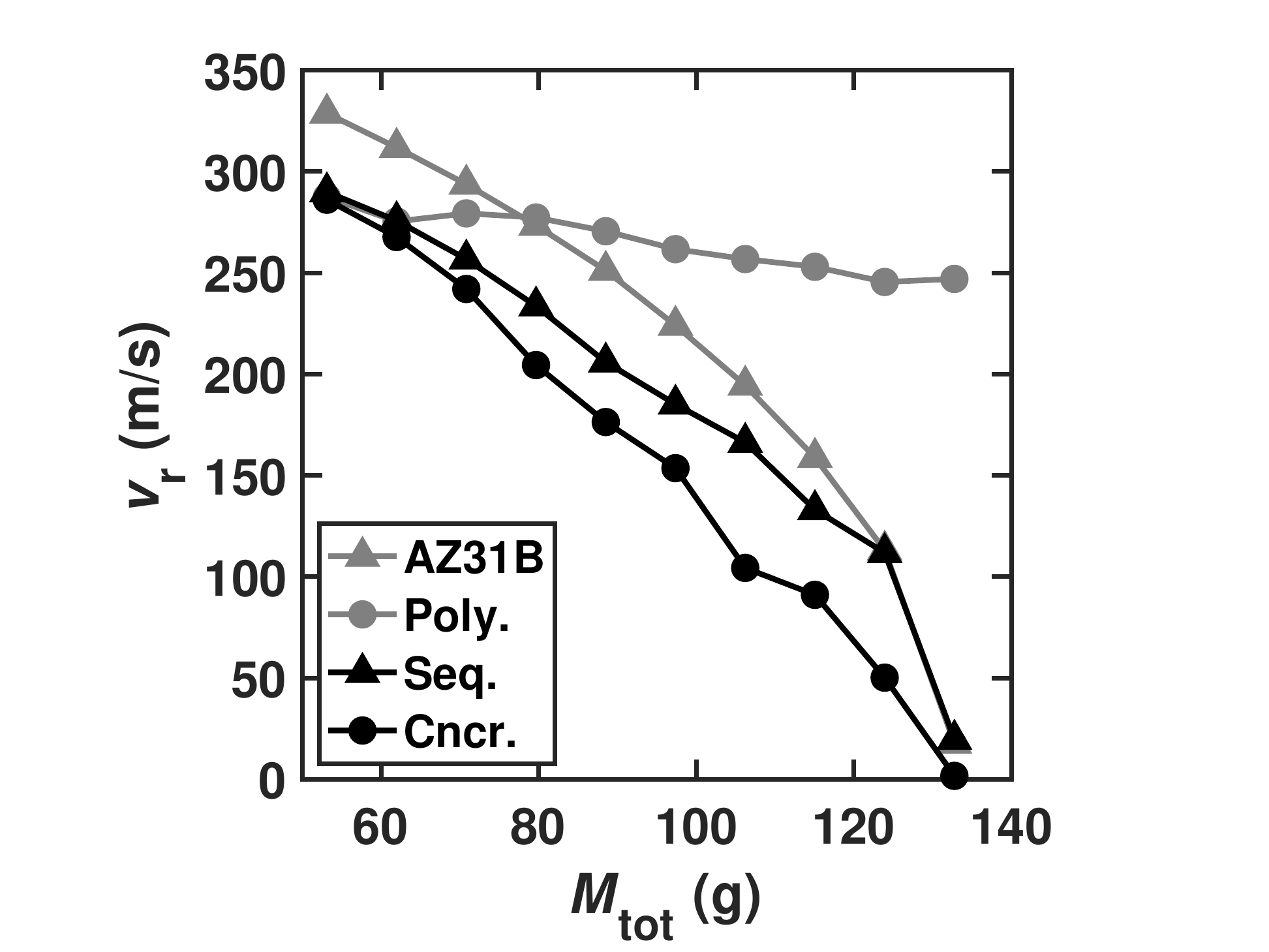}}
\subfloat[ ]{\includegraphics[width=1.95in]{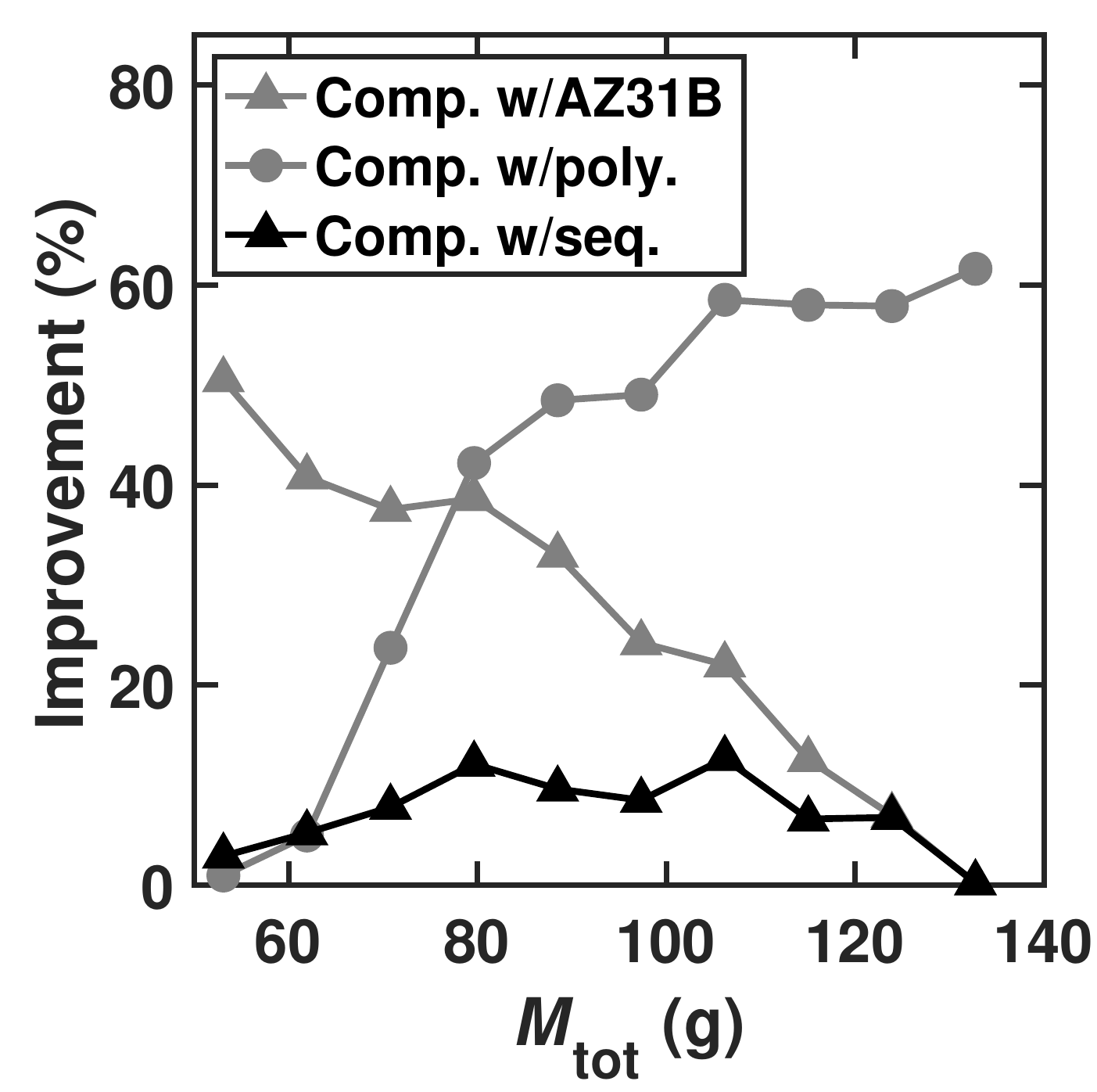}}
\caption{Optimal design and performance for various total mass of the double-layer plate using both the sequential (seq.) and concurrent (cncr.) design strategies. (a) The optimal material properties. (b) The optimal plate thicknesses. (c) Residual velocity for the monolithic and double-layer plates. (d) Improvement of energy dissipation using the concurrent design strategy over the monolithic plates and sequential design strategy.}
\label{fig:param}
\end{figure}

We repeat the optimal design calculations for various values of the total mass of the plate over the range of $53.1$ to $132.75$~g. The results are collected in Fig.~\ref{fig:param}. Fig.~\ref{fig:param}(a) shows how the optimal mechanical properties of AZ31B and polyurea change with the total mass of the plate. The optimal properties of AZ31B are concentrated near the point with the greatest strength, i.~e., the maximum value of $A$, for both design strategies, though the concurrent design selects properties that are slightly stronger and less ductile than the sequential design. By contrast, the properties of polyurea selected by the concurrent design are distributed widely as a functions of total mass, whereas the sequential design properties concentrate around the point of maximum strength.

Fig.~\ref{fig:param}(b) shows the optimal thickness of AZ31B and polyurea layers. In both strategies, the thickness $t^1$ of the AZ31B increases, while the thickness $t^2$ of polyurea decreases, with total mass. In addition, the thickness of AZ31B tends to zero with small total mass while that of polyurea tends to zero with large total mass. Thus, monolithic plates yield the best  performance when the allowable mass is either small or large, whereas the double-layer plates are best in the intermediate range. The intermediate range is larger for the concurrent design strategy, showing that the sequential and concurrent designs may differ qualitatively as well as quantitatively. We also observe a plateau in the thickness of the polyurea at intermediate values of the mass in both design strategies, which effectively uses the extra mass to increase the thickness of the AZ31B layer.

Finally, the performance of the various plates in terms of residual velocity and energy dissipation is shown in Fig.~\ref{fig:param}(c) and Fig.~\ref{fig:param}(d), respectively. We see from the figures that the residual velocity decreases with increasing mass, with the monolithic polyurea plate providing optimal performance at small mass and the monolithic AZ31B at large mass. The double-layer plates perform better at intermediate mass, with the concurrent design delivering best performance. In terms of energy dissipation, the concurrent design improves the ballistic performance of the plate by more than $53\%$ over monolithic AZ31B, $61\%$ over monolithic polyurea and $12\%$ relative to the sequential design.

\section{Summary and concluding remarks} \label{sec:summ}

We have presented and assessed two approaches, concurrent and sequential, for determining the optimal mechanical and structural designs of double-layered armor plates under high-speed impact. The sequential design strategy first determines the optimal mechanical properties of each material for monolithic plates with fixed structural properties. Then, the  composite layered plate is optimized while holding fixed at the optimal material properties obtained from the previous step. Contrariwise, the concurrent and goal-oriented design strategy optimizes the ballistic performance of the double-layered plate over both the mechanical and structural properties simultaneously. We have applied these strategies to the design of a double-layered plate against ballistic impact using hard AZ31B as the front layer and soft polyurea as the back layer. We have developed a non-intrusive, high-performance computational framework based on DAKOTA Version $6.12$ and GMSH Version $4.5.4$, which provide greater portability and broader usage. For the impact  calculations, we use the commercial finite-element package LS-DYNA.

Several significant findings afforded by the calculations are noteworthy. Both the double-layered plates designed by the sequential and concurrent strategies are capable of significantly improving the ballistic performance in comparison with the plates consisting of their monolithic counterparts when the total mass of the plates is fixed in an appropriate range. However, the concurrently-designed plate exhibits superior performance relative to the sequentially-designed one. Using the total dissipated energy as a reference, the concurrent design can improve the ballistic performance of the plate by more than $53\%$ with respect to the monolithic AZ31B, $61\%$ with respect to monolithic polyurea and $12\%$ over the sequential design. Moreover, the material failure mechanisms involved in the impact tests bear emphasis, including petaling, plugging and fragmentation for the monolithic AZ31B plate, localized tensile drawing for the monolithic polyurea plate, and delamination and plugging for the double-layered plates. The advantage of the concurrent design strategy over the sequential results from differences in the dominant failure mechanisms mainly controlled by the mechanical properties of polyurea.

We close with a discussion of various ways in which the concurrent goal-oriented approach can be generalized. Whereas we have studied a two-material system, the methodology is applicable to an arbitrary number of materials. Similarly, the geometric complexity and the number of material and geometric parameters can easily be generalized. Finally, the methodology can also be combined with topology optimization. In this context, it is common to use SIMP or some other interpolation scheme~\cite{bendsoe2013topology} to interpolate between two materials, or a material and a void. Evidently, this type of interpolation can be generalized to material properties that range over entire domains, instead of having fixed values. These and other generalizations suggest worthwhile directions for further research.

%

\section*{Acknowledgement}

This research was sponsored by the Army Research Laboratory and was accomplished under Cooperative Agreement Number W911NF-12-2-0022. The views and conclusions contained in this document are those of the authors and should not be interpreted as representing the official policies, either expressed or implied, of the Army Research Laboratory or the U.S. Government. The U.S. Government is authorized to reproduce and distribute reprints for Government purposes notwithstanding any copyright notation herein.

\bibliographystyle{unsrt}
\bibliography{MaterialsByDesign}

\end{document}